\documentclass{azh}

\usepackage{graphicx}
\usepackage{natbib}
\usepackage{lscape}

\def\degr{\hbox{$^\circ$}}
\def\arcmin{\hbox{$^\prime$}}
\def\arcsec{\hbox{$^{\prime\prime}$}}

\begin{document} 

\title{Young Stellar Complexes in the Giant Galaxy UGC 11973}

\author{A.~S.~Gusev,$^1$ 
        F.~H.~Sakhibov,$^2$
        and O.~V.~Ezhkova$^1$}

\institute{$^1$ Sternberg Astronomical Institute, Lomonosov Moscow State University, 
                Moscow, Russia \\
           $^2$ Applied Sciences University Hessen-Friedberg, Friedberg, Germany}

\date{Received December 11, 2019; revised January 24, 2020; accepted 
      January 24, 2020}
\offprints{Alexander~S.~Gusev, \email{gusev@sai.msu.ru}}

\titlerunning{Young Stellar Complexes}
\authorrunning{Gusev et al.} 

\abstract{We present results of the analysis of photometric and spectroscopic 
observations of the young stellar complexes in the late giant spiral galaxy UGC 11973. 
Photometric analysis in the $UBVRI$ bands have been carried out for the 13 largest 
complexes. For one of them, metallicity of the surrounding gas $Z=0.013\pm0.005$, 
the mass $M=(4.6\pm1.6)\cdot10^6 M_{\odot}$, and the age of the stellar complex 
$t=(2.0\pm1.1)\cdot10^6$~yr were evaluated, using spectroscopic data. It is shown 
that all complexes are massive ($M\ge1.7\cdot10^5 M_{\odot}$) stellar groups
younger than $3\cdot10^8$~yr. \\

{\bf DOI:} 10.1134/S1063772920060025 \\

Keywords: {\it galaxies: spiral -- galaxies: star formation -- galaxies: star 
complexes}
}

\maketitle

\section{INTRODUCTION}

\begin{figure*}
\vspace{1mm}
\centerline{\includegraphics[width=16cm]{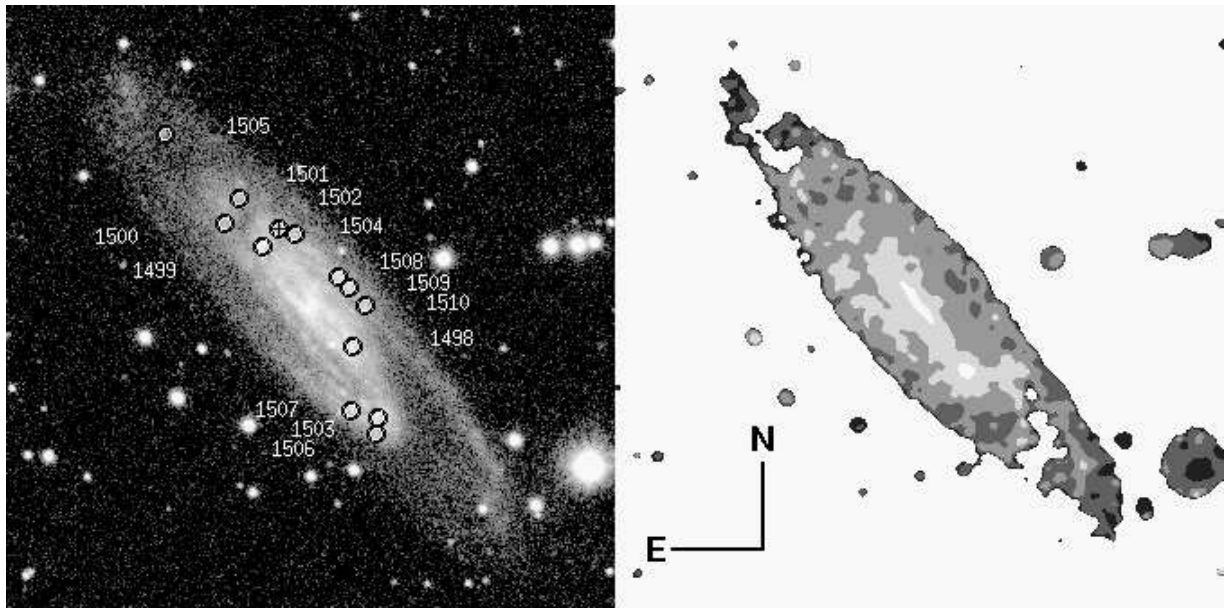}}
\caption{The image of UGC 11973 in the $B$ band (left) and the map of the color 
index $(B-V)_0^i$ corrected for absorption in the galaxy and for the inclination 
of the disk of UGC 11973 (right). In the image on the left, the circles indicate 
star formation complexes, a circle with a cross (No. 1502) indicates the complex 
studied spectroscopically, the numbers indicate the numbers of regions according 
to the catalog \citet{gusev2018}. To the right, the areas with the color indices 
$>0.8$, 0.6--0.8, 0.4--0.6, 0.2--0.4 и $<0.2^m$ are highlighted; redder areas are 
shown by lighter color. The size of images is $3.0\arcmin \times 3.0\arcmin$.
\label{figure:fig1}}
\end{figure*}

Among relatively close galaxies, UGC 11973 is one of the brightest, largest, 
and most massive stellar systems (see Table~\ref{table:gal}). Its radius 
reaches 30 kpc and absolute magnitude $M(B)_0^i < -22^m$. The analysis of 
the rotation curve has shown that the total mass of the galaxy within only 
10 kpc from the center (1/3 of the radius $D_{25}$) is 
$2.4\cdot10^{11} M_{\odot}$ \citep{afanasiev1988}. The total stellar mass 
of UGC 11973, as estimated from the luminosity in the $B$ and $K$ bands, 
is $9.4\cdot10^{10} M_{\odot}$ \citep{graur2017}. According to observational 
data, in the 21 cm line, the HI mass is $2\cdot10^{10} M_{\odot}$ 
\citep{courtois2015}. The galaxy belongs to a small group \cite{garcia1993}, 
but has no nearby satellites. Despite the large inclination ($i=81\degr$), 
symmetric structure of UGC 11973 is clearly traced: powerful spiral arms 
with dust lines and a weak bar (Fig.~\ref{figure:fig1}).

Additionally to the large inclination, the difficulty in the study of this 
galaxy is associated with its proximity to the plane of the Milky Way, 
because of which the total attenuation of the light in the $B$ band exceeds 
$1.5^m$. Apparently, this explains the relatively poor knowledge about 
UGC 11973. Although detailed studies of the galaxy were not carried out 
\citep[except in ][]{afanasiev1988}, it was observed in large projects 
in a wide wavelength range, from UV to radio. Observations in the radio 
\citep{marvil2015,condon2002}, far and near IR (IRAS \citep{strauss1992} 
and 2MASS projects \citep{skrutskie2006}) and optical ranges 
\citep{gusev2015}, as well as a significant brightness of the galaxy 
in the UV range (GALEX 
project\footnote{http://galex.stsci.edu/GR6/?page=explore\&objid= \\
6372147238576064570}), 
consistently point to the active, constant in time star formation, 
characteristic for the massive late-type spiral galaxies. The color 
indices of UGC 11973, corrected for the Galactic absorption and 
absorption due to the inclination of the disk, decrease from 
$(U-B)_0^i = 0.38\pm0.03^m$ and $(B-V)_0^i = 0.85\pm0.04^m$ in the 
central part (nucleus and bulge) to $(U-B)_0^i = 0.08\pm0.19^m$, 
$(B-V)_0^i = 0.59\pm0.13^m$ in the region of spiral arms outside the 
bright regions of star formation \citep{gusev2015} 
(Fig.~\ref{figure:fig1}).

\begin{table*}
\begin{center}
\caption{The main parameters of UGC 11973.}
\label{table:gal}
\begin{tabular}{|c|c||c|c||c|c||c|c|}
\hline\hline
Type       & SAB(s)bc (3.9) & $i$  & $81\degr$ & $D_{25}$      & $3.46\arcmin$   
& $V_{\rm max}$ & $231\pm7$ km/s \\
$m(B)$     & $13.34^m$      & P.A. & $39\degr$ & $D_{25}$      & 59.2 kpc        
& $A(B)_{\rm G}$ & $0.748^m$ \\
$M(B)_0^i$ & $-22.47^m$     & $d$  & 58.8 Mpc  & $V_{\rm rad}$ & $4215\pm8$ km/s 
& $A(B)_i$      & $0.85^m$ \\[1mm]
\hline
\end{tabular}
\end{center}
\end{table*}

The basic information about the galaxy -- morphological type, 
apparent magnitude $m(B)$, corrected for the Galactic absorption 
and the absorption due to the inclination of the disk absolute stellar 
magnitude $M(B)_0^i$, the inclination $i$ , and position angle P.A. 
of the disk, the distance $d$, diameter along the isophote $25^m$ in 
the $B$ band with the account of the galactic absorption and 
absorption caused by the inclination of the galaxy $D_{25}$, radial 
velocity $V_{\rm rad}$, maximum rotation velocity $V_{\rm max}$, 
galactic absorption $A(B)_{\rm G}$ and absorption caused by the 
inclination of the galaxy $A(B)_i$ -- are presented in 
Table~\ref{table:gal}. The data on the magnitude of absorption 
in our Galaxy ($A(B)_{\rm G}$) and the morphological type of 
UGC 11973 are given according to the 
NED \footnote{http://ned.ipac.caltech.edu/} database, the 
remaining parameters were taken from the 
HyperLEDA\footnote{http://leda.univ-lyon1.fr/} database. Note that 
in a number of papers the distance to the galaxy is taken as 
49–55 Mpc, obtained by the Tully–Fisher method in \citet{sorce2014}. 
This somewhat reduces the estimates of the luminosity, mass, and linear 
size of UGC 11973. In particular, the latest version of the HyperLEDA 
provides the absolute magnitude of the galaxy 
$M(B)_0^i = -22.28\pm0.35^m$. In this study, we use the values given 
in Table~\ref{table:gal} based on the measurement of radial velocity 
of UGC 11973 \citep{afanasiev1988}.

The aim of this study is to analyze the photometric parameters and to 
estimate the physical ones of the 13 largest star formation complexes 
found in the galaxy. The basis for the analysis is the $UBVRI$ 
photometric data of the complexes that we obtained earlier, as well as the 
results of spectroscopic observations of complex No.~1502 
\citep{gusev2018}, which were not published previously.

This work is a part of an extensive project of the study of the physical 
parameters of the stellar population of the regions of current star 
formation in the disks of galaxies, based on the complex spectral and 
photometric observations \citep{gusev2012,gusev2013,gusev2016,gusev2018}. 
The results of spectroscopic studies were published by us earlier in 
\citet{gusev2012,gusev2013} with a single exception -- the HII region 
(complex No.~1502) in the galaxy UGC 11973. The reason for the 
''omission'' was that for this HII region we were not able to measure the 
oxygen lines [O\,{\sc iii}]$\lambda$4959 and $\lambda$5007 (see 
Section 3.1 for more details). This required a special nonstandard method 
for determination of the chemical composition of the gas in the region. 
Such a technique was developed only in 2016 \citep{pilyugin2016}.

In the next work within the framework of the project, we plan to study 
physical parameters in several hundred star formation regions using, among 
other things, spectral data from the literature. For these regions, 
chemical composition of the gas will be determined by standard methods 
using the intensities of the oxygen lines. The nonstandard method for 
determination of the chemical parameters of the gas in the HII region in 
the galaxy UGC 11973 is, in our opinion, of independent interest.

\section{OBSERVATIONS AND THEIR PROCESSING}

Photometric observations of the galaxy in filters $UBVRI$ carried out 
at the 1.5-meter telescope of the Maidanak Observatory (Uzbekistan) 
were described by us earlier \citep{gusev2015}. Identification of star 
formation complexes that was carried out using the 
SExtractor~5\footnote{http://sextractor.sourceforge.net/} program 
with the detection threshold above the local background equal to 
$5\sigma$ and the number of the pixels $\ge10$ above the threshold, 
and the photometry of 13 detected complexes was described in 
\citet{gusev2018}. The catalog of photometric parameters of star-forming 
regions is also presented in the electronic 
form\footnote{http://lnfm1.sai.msu.ru/$\sim$gusev/sfr\_cat.html}.                  
                                                            
Spectral observations of one of the brightest HII regions in UGC 11973 
-- the complex No.~1502, located $23.0\arcsec$ to the north and 
$9.5\arcsec$ east of the center of the galaxy (see 
Fig.~\ref{figure:fig1}) -- were carried out by one of the authors of 
this study on August 23/24, 2006 at 6-meter telescope BTA of Special 
Astrophysical observatory of the Russian Academy of Sciences using 
the SCORPIO focal reducer in the multi-slit mode 
\citep[for a detailed description of the device see][]{afanasiev2005}. 
An EEV 42-40 CCD camera was used as a receiver. The size of the matrix 
is $2048 \times 2048$ pixels, which provides the field of view $6/arcmin$ 
for the image scale $0.178\arcsec$/pixel. In multi-slit mode, the size 
of the slits is $1.5\arcsec \times 18\arcsec$. Observations were 
carried out at air mass $M(z)=1.16-1.33$. The seeing was $1.8\arcsec$. 
In total, four expositions of 900 sec each were made. After each exposition, 
the positions of the slits were shifted left and right along the slit in 
the increments of 30 pixels.

To carry out standard data processing and calibration, before and after 
observations of the galaxy, the images of bias, the ''flat field'', the 
spectra of the helium-neon-argon lamp, and the comparison star 
BD+25$\degr$4655 from the catalog of spectrophotometric standards of 
\citet{oke1990} were obtained.

Further processing was carried out according to the standard procedure 
using the ESO-MIDAS image processing system. The main stages of the 
processing included: elimination of the traces of cosmic rays, 
determination and correction of the data for the offset of the matrix 
amplifier (bias) and ''flat field'', transformation into the wavelength 
scale using the spectrum of He-Ne-Ar lamp, background subtraction, 
transformation of the instrumental fluxes into absolute ones according 
to the observations of the standard star, correction for atmospheric 
absorption, integration of two-dimensional spectra in the selected 
apertures and obtaining one-dimensional spectra of the HII region, 
and addition of spectra.

\begin{table*}
\begin{center}
\caption{Spectral parameters of the HII region, physical and chemical parameters 
of the gas in the complex No. 1502.}
\label{table:spec}
\begin{tabular}{|c|c||c|c||c|c|}
\hline\hline
Line & Flux $F$, & Parameter & Value & Parameter & Value \\
 & 10$^{-16}$erg/(s$\cdot$cm$^{2}$) & & & & \\
\hline
H$\beta$            &  $4.27\pm0.49$ & $c$(H$\beta$) & $1.27\pm0.26$ & & \\
$[$N\,{\sc ii}]6548 &  $6.16\pm1.06$ & [N\,{\sc ii}]6548+6584 & $2.12\pm0.76$ 
& 12+(O/H) & $8.67\pm0.18$ \\
H$\alpha$           & $48.24\pm1.11$ & [S\,{\sc ii}]6717+6731 & $0.99\pm0.37$ 
& $Z$ & $0.013\pm0.005$ \\
$[$N\,{\sc ii}]6584 & $17.97\pm1.07$ & [S\,{\sc ii}]6717/[S\,{\sc ii}]6731 & $1.24\pm0.42$ 
& $T_e$, K & $\sim5000$ \\
$[$S\,{\sc ii}]6717 &  $6.65\pm1.01$ & EW(H$\alpha$), \AA \, & $26.84\pm1.24$ 
& $n_e$, cm$^{-3}$ & $\le300$ \\
$[$S\,{\sc ii}]6731 &  $5.37\pm1.01$ & EW(H$\beta$), \AA \,  &  $3.26\pm0.49$ & & \\[1mm]
\hline
\end{tabular}
\end{center}
\end{table*}

A detailed description of the methodology for the processing spectral 
observations was presented by us in \citet{gusev2012,gusev2013}.
                                                            
The equivalent widths of the emission lines H$\alpha$ and H$\beta$ 
were estimated from the spectrum of the HII region with the allowance 
for the continuum. Such a spectrum was constructed by subtracting 
the spectrum of the surrounding galactic disk substrate from the 
spectrum of the HII region. This made it possible to exclude the 
contribution of the stars and gas from the galactic disk to the 
radiation coming from the HII region.

Observed fluxes in the lines are presented in the left column of 
Table~\ref{table:spec}. The middle column of the table shows the 
absorption coefficient $c$(H$\beta$), corrected for interstellar 
light absorption relative intensities of the lines 
[N\,{\sc ii}]$\lambda$6548+$\lambda$6584 and 
[S\,{\sc ii}]$\lambda$6717+$\lambda$6731 (in the units of $I$(H$\beta$)), 
the ratio of the sulfur lines 
[S\,{\sc ii}]$\lambda$6717/[S\,{\sc ii}]$\lambda$6731, equivalent line 
widths of H$\alpha$ and H$\beta$ lines.

Absorption of the gas emission lines was taken into account based on the 
Balmer decrement using the theoretical ratio H$\alpha$/H$\beta$ 
\citep{osterbrock1989} for the case B of recombination at the electron 
temperature $10^4$K and analytical approximation of Whitford's 
interstellar reddening law from \citet{izotov1994}. Moreover, the 
equivalent width of the hydrogen absorption lines EW$_a$($\lambda$) was 
taken as 2\AA, which is the average value for the HII regions 
\citep{mccall1985}. For the lines of other chemical elements, the value 
of EW$_a$($\lambda$) was taken as equal to 0.

\section{THE ANALYSIS OF RESULTS}

\subsection{Physical and chemical parameters of the gas in the star 
formation complex No. 1502}

A number of gas characteristics can be estimated using gas emission lines. 
The [O\,{\sc iii}]$\lambda$4959 and $\lambda$5007 lines are very 
important for this. However, as was mentioned already in the Introduction, 
due to the relative weakness of these lines and large absorption in 
this region (see Table~\ref{table:spec}), it appeared impossible to 
measure them. Therefore, we carry out the analysis that is possible on 
the basis of the measurements of the obtained lines of hydrogen, 
nitrogen, and sulfur.         
                                                          
The classic diagrams for classification of the objects of various types 
of emission lines ($BPT$ diagrams 
$\log$([(N\,{\sc ii}]$\lambda$6584/H$\alpha$) -- 
$\log$([O\,{\sc iii}]$\lambda$5007/H$\beta$) and 
$\log$([S\,{\sc ii}]$\lambda$6717+$\lambda$6731)/H$\alpha$) -- 
$\log$([O\,{\sc iii}]$\lambda$5007/H$\beta$)) allow us to determine 
the main mechanism of the gas excitation in the emission region. 
Despite the lack of the measurements of the 
[O\,{\sc iii}]$\lambda$5007 line, we can estimate the upper limit of 
its intensity -- $\log$([O\,{\sc iii}]$\lambda$5007/H$\beta))<0$. 
Given the relative intensities of the nitrogen and sulfur lines 
$\log$([(N\,{\sc ii}]$\lambda$6584/H$\alpha$)$=-0.43\pm0.04$ and 
$\log$([S\,{\sc ii}]$\lambda$6717+$\lambda$6731)/H$\alpha$)$=-0.60\pm0.03$, 
the source studied by us is located in the $BPT$ diagrams in the 
region occupied by the HII regions, that is, the objects with 
photoionization \citep[according to the models of][]{kewley2001}.

To assess the chemical composition and physical characteristics of 
the gas surrounding the stellar complex in the absence of data on 
the intensities of the oxygen lines, we can use the empirical 
dependences and correlations obtained by \citet{pilyugin2016}.

Relative intensity of the nitrogen lines indicates high metallicity 
of the HII region; therefore, we can use formula (9) from 
\citet{pilyugin2016} to estimate the oxygen abundance: 
$12+{\rm (O/H)}=8.67\pm0.18$ or $Z=0.013\pm0.005$, which, 
within the errors, corresponds to the solar metallicity. According 
to the relationship between the relative intensity of the nitrogen 
line $\log$($I$([(N\,{\sc ii}]$\lambda$6548+$\lambda$6584)/$I$(H$\beta$)) 
and the electron temperature $T_e$ from \citet{pilyugin2016}, we 
estimated the gas temperature $T_e\sim5000{\rm K}$.                                   

Such characteristics -- high metallicity, close to the solar one 
and relatively low electron temperature -- are typical for HII regions 
in the giant spiral galaxies. Similar O/H and $T_e$ values were 
obtained by us also for the large nearby galaxy NGC 6946 
\citep{gusev2013}, where an anticorrelation between the values of 
O/H and electron temperature was also noted.

The ratio of sulfur lines 
[S\,{\sc ii}]$\lambda$6717/[S\,{\sc ii}]$\lambda$6731$=1.24\pm0.42\ge1$ 
from the HII region corresponds to the electron density of gas 
$n_e \le 300$~cm$^{-3}$.

Similar relatively low densities are characteristic for the giant 
HII regions observed in other galaxies 
\citep[see, for example,][]{kennicutt1984,gutierrez2010}. Note that 
the diameter of region No. 1502 was estimated by us as 550 pc 
\citep{gusev2018}, which is typical scale for stellar complexes -- 
the largest regions of coherent star formation \citep{efremov1995}.

Our estimates of the chemical and physical parameters of the gas in 
the region of star formation No. 1502 are presented in the right column 
of Table~\ref{table:spec}.

\subsection{Photometric and Physical Parameters of the Star Formation 
Complexes in UGC 11973}

The study of the earliest life stages of star formation regions and 
evaluation of their physical parameters is a difficult task, due to 
the action of gas and dust. Perhaps the most difficult task is 
evaluation of the age of the stellar population. While, in the 
closest galaxies, we can resolve a young stellar group into separate 
stars and to determine their age by the color-magnitude diagram 
\citep[see, for instance,][]{whitmore2011}, for the more distant 
galaxies, spectroscopic or photometric data or a combination of them 
is used. The spectroscopic method includes both an estimate of the 
spectral age indices (for instance, equivalent widths EW(H$\alpha$) 
and EW(H$\beta$), the ratio [O\,{\sc iii}]/H$\beta$, the fluxes in 
the He\,{\sc ii} emission lines, etc.) and a direct comparison of 
the observed spectra with the model ones 
\citep{copetti1986,stasinska1996,schaerer1998,bastian2005,bastian2006,
bastian2009,konstantopoulos2009,wofford2011}. The photometric 
method includes comparison of the multicolor photometry data for 
star formation regions with predictions of the evolutionary or 
population synthesis models \citep{searle1980,elson1985,
bresolin1996,chandar2010,hollyhead2015,hollyhead2016}.

\begin{table*}
\begin{center}
\caption{Photometric and physical characteristics of starburst complexes.}
\label{table:data}
\begin{tabular}{|c|c|c|c|c|c|}
\hline\hline
No.  & $M(B)$, mag  & $U-B$ & $B-V$ & $V-R$ & $V-I$ \\
 & $d$, pc & $r$, kpc & $I_{\rm gas}/I_{\rm total}(B)$, \% & $t$, Myr & 
$m$, $10^6 M_{\odot}$ \\
\hline
1498 & --13.29$\pm$0.15 & --0.27$\pm$0.20 & 0.21$\pm$0.13 & 0.08$\pm$0.18 & 0.34$\pm$0.25 \\
 & 700 &  6.01 & -- & $\le300$ & $\ge0.21$ \\
1499 & --15.29$\pm$0.04 & --0.50$\pm$0.08 & 0.18$\pm$0.08 & 0.26$\pm$0.10 & 0.67$\pm$0.10 \\
 & 850 &  6.48 & -- & $\le45$ & $\ge1.4$ \\
1500 & --13.56$\pm$0.12 & --0.37$\pm$0.30 & 0.23$\pm$0.16 & 0.05$\pm$0.21 & 0.45$\pm$0.22 \\
 & 650 & 12.96 & -- & $\le300$ & $\ge0.27$ \\
1501 & --13.22$\pm$0.16 & --0.73$\pm$0.28 & 0.19$\pm$0.21 & 0.06$\pm$0.24  & 0.71$\pm$0.16 \\
 & 550 & 12.98 & -- & $\le47$ & $\ge0.19$ \\
1502 & {\bf --15.99$\pm$0.56} & {\bf --1.19$\pm$0.23} & {\bf --0.34$\pm$0.27} 
& {\bf --0.06$\pm$0.27} & {\bf 0.19$\pm$0.35} \\
 & 550 & 14.53 & {\bf 7} & {\bf 2.0$\pm$1.1} & {\bf 4.6$\pm$1.6} \\
1503 & --13.55$\pm$0.06 & --0.62$\pm$0.11 & 0.25$\pm$0.08 & 0.41$\pm$0.10 & 0.70$\pm$0.09 \\
 & 650 & 14.91 & -- & $\le25$ & $\ge0.29$ \\
1504 & --13.30$\pm$0.14 & --0.61$\pm$0.19 & 0.27$\pm$0.13 & -- & 0.85$\pm$0.15 \\
 & 600 & 18.35 & -- & $\le32$ & $\ge0.21$ \\
1505 & --14.06$\pm$0.04 & --1.29$\pm$0.04 & 0.11$\pm$0.06 & 0.14$\pm$0.09 & 0.47$\pm$0.09 \\
 & 550 & 19.35 & -- & $\le5.1$ & $\ge0.45$ \\
1506 & --13.05$\pm$0.06 & --0.78$\pm$0.18 & --0.07$\pm$0.11 & 0.26$\pm$0.15 & --0.11$\pm$0.36 \\
 & 550 & 20.55 & -- & $\le4.7$ & $\ge0.17$ \\
1507 & --14.42$\pm$0.04 & --0.75$\pm$0.09 & 0.14$\pm$0.06 & 0.33$\pm$0.07 & 0.30$\pm$0.08 \\
 & 650 & 20.74 & -- & $\le5.3$ & $\ge0.63$ \\
1508 & --14.35$\pm$0.07 & --0.56$\pm$0.13 & 0.05$\pm$0.14 & 0.27$\pm$0.14 & 0.24$\pm$0.18 \\
 & 600 & 21.26 & -- & $\le53$ & $\ge0.58$ \\
1509 & --14.43$\pm$0.06 & --0.57$\pm$0.12 & --0.10$\pm$0.12 & 0.18$\pm$0.13 & 0.30$\pm$0.19 \\
 & 500 & 22.43 & -- & $\le4.7$ & $\ge0.57$ \\
1510 & --14.43$\pm$0.06 & --0.67$\pm$0.10 & 0.09$\pm$0.08 & 0.13$\pm$0.09 & 0.55$\pm$0.08 \\
 & 650 & 22.68 & -- & $\le5.1$ & $\ge0.63$ \\[1mm]
\hline
\end{tabular}
\end{center}
\end{table*}

\begin{figure*}
\vspace{9mm}
\centerline{\includegraphics[width=13.6cm]{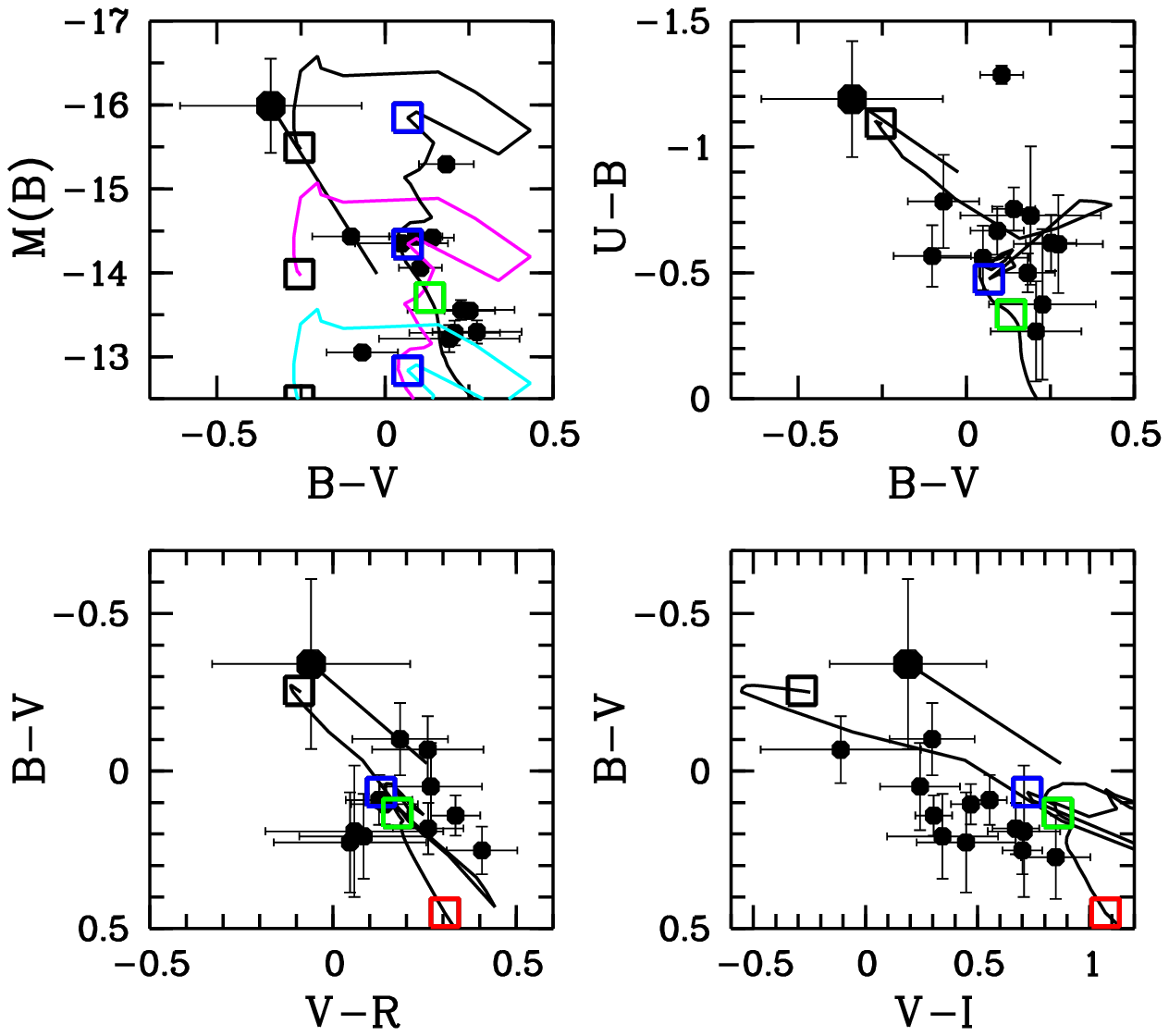}}
\caption{Color-magnitude diagram $(B-V) - M(B)$ and two-color diagrams 
$(U-B)-(B-V)$, $(B-V)-(V-R)$, and $(B-V)-(V-I)$ for star formation complexes in 
the galaxy. For the complex No. 1502 (large black circle), the absolute stellar 
magnitude $M(B)_{\rm c}$ and color indices corrected for the light absorption and 
the contribution of the gas emission lines are shown; for the remaining 
star-formation complexes (small black circles) luminosities $M(B)_0^i$ and color 
indices corrected for absorption in the galaxy and absorption associated with the 
inclination of the UGC 11973 disk are shown. Measurement errors are indicated. 
Thick black lines show the displacement of the objects in the diagrams along the 
absorption line. The length of the segments corresponds to the absorption value 
$A(B)=1.99^m$, equal to the difference between the absorption 
$A(B)_{\rm c}=3.59^m$ in the complex No. 1502, determined by the Balmer decrement, 
and the sum of the absorption $A(B)_{\rm G}+A(B)_i$ (see Table~\ref{table:gal}). 
Thick lines are evolutionary tracks of stellar systems with metallicity $Z=0.012$. 
In the color-magnitude diagram the tracks of stellar systems with masses 
$4\cdot10^6 M_{\odot}$ (black color), $1\cdot10^6 M_{\odot}$ (purple color), and 
$2.5\cdot10^5 M_{\odot}$ (blue color) are shown. The squares are the positions 
of stellar systems with the age 1~Myr (black), 10~Myr (blue), 100~Myr (green), and 
1~Gyr (red), respectively.
\label{figure:fig2}}
\end{figure*}

Although the ages estimated for the same stellar clusters using 
spectroscopic and photometric data are in fairly good agreement 
\citep{searle1980,whitmore2011,wofford2011}, \citet{kim2012}, 
who studied individual clusters in M83, which can be resolved 
into stars, found that the correlation between the ages of stellar 
groups obtained from the ages of individual stars in the region 
and the ages obtained from integral color indices using the 
standard photometric method is not very strong.

Spectroscopic methods usually provide high-quality estimates of 
the age, but the latter can be determined for a limited number 
of objects. The main problem in estimating the photometric age 
is taking into account the influence of gas and dust in the 
measured photometric fluxes. The absence of independent data 
on the chemical composition and absorption leads to the 
degeneracy of the ''age--metallicity'' and 
''age--absorption'' diagrams in the comparative analysis with 
theoretical evolutionary models of stellar clusters 
\citep{scalo1986}.

To assess the physical parameters of the stellar population in 
the star formation complexes, we used the technique described in 
detail in \citet{gusev2007a,gusev2016} and tested in 
\citet{gusev2007b} and \citet{gusev2016}. This is based on the 
observed luminosities and color indices of the objects obtained 
from the photometry, and the intensities of the emission lines, 
the estimates of metallicity and absorption in the gas, obtained 
from the spectroscopic observations. Then, using evolutionary 
models with specific chemical composition, from the luminosity 
and color indices of star formation regions corrected for 
the light absorption and the contribution of emission lines, 
we can estimate the mass and age of the young stellar 
population. When this technique is applied for solving the 
problem of determination of the parameters of mass $m$ and age 
$t$, not only all local minima of the deviation functional 
are searched, but their depth is also calculated. As 
solution of the problem, the deepest minimum is taken.                                           

In the simulation, version~3.1 of the Padova isochrones 
grid was used \citep[see, for example,][]{marigo2017}, 
which is accessible via the 
CMD\footnote{http://stev.oapd.inaf.it/cgi-bin/cmd} online 
server. The sets of stellar evolutionary tracks for this 
version were calculated for the Salpeter initial mass function 
and the mass range from $0.15M_{\odot}$ to $100M_{\odot}$.

To assess the age, a single starburst model (SSP model) was 
used. Although in large stellar complexes consisting of star 
clusters systems and OB-associations, star formation can 
occur over a longer period of time, the choice of star 
formation mode for them is ambiguous. Continuous 
constant-rate star formation in such systems is an extreme case. 
The most probable seems to be a series of star formation bursts 
that have various power and gaps between them. The largest 
contribution to the color characteristics of such a complex will 
be made by the last major burst of star formation. Due to the 
uncertainty in the history of star formation in large stellar 
complexes, we decided to abandon the modeling of the constant 
star formation mode. At the same time, it is worth remembering 
that the ages of the young stellar complexes that we determine 
are ''photometric'', but not real physical ages.

Relative contribution of the gas continuum to the radiation 
in wide photometric bands ($I_{\rm gas}/I_{\rm total}$) was 
estimated using equations for the spectral intensity of 
radiation near the boundaries of the hydrogen series, 
free-free emission, and two-photon emission 
\citep{osterbrock1989,kaplan1979}. The contribution of the 
emission lines was calculated by summation of the intensities 
of the lines that appear in the given photometric band. The 
fluxes for unmeasured emission lines were calculated from the 
estimates of the emission measure EM, using the equations from 
\citet{kaplan1979} and \citet{osterbrock1989}. In total, 18 main 
lines of the interstellar medium were taken into account.

In the evaluation of the ages and masses, we used color indices 
$U-B$ and $B-V$, since in the case of a young stellar population, 
the fluxes in the bands $R$ and $I$ have weak sensitivity to 
the changes of the age, and their actual measurement errors 
increase the uncertainties of the estimates of the age and mass.

The value of luminosity in the $B$ band, $M(B)_{\rm c}$, color 
indices $(U-B)_{\rm c}$, $(B-V)_{\rm c}$, $(V-R)_{\rm c}$ and 
$(V-I)_{\rm c}$ corrected for the light absorption and the 
contribution of the gas emission lines to the total flux, the 
estimates of the gas contribution $I_{\rm gas}/I_{\rm total}$, 
the age $t$ and mass $m$ obtained for the complex No. 1502, are 
presented in Table~\ref{table:data} and are shown there in 
boldface. For the rest of the star-forming complexes in the 
galaxy, we list in the table the values $M(B)_0^i$, $(U-B)_0^i$, 
$(B-V)_0^i$, $(V-R)_0^i$ and $(V-I)_0^i$, corrected for the 
absorption in the galaxy and the absorption caused by the 
inclination of the UGC 11973 disk. Under the assumption that 
the absorption in the HII regions is greater than or equal to 
the sum of the absorption $A_{\rm G}$ and $A_i$ and the 
metallicity of the stellar population in them corresponds to 
the solar one, we estimated the lower limit of the mass and 
the upper limit of the age for the remaining complexes in the 
galaxy.

The limiting values of the mass and age correspond to the case 
of the absence of additional absorption in the HII region 
(absorption is equal to $A_{\rm G}+A_i$). In the case of 
additional absorption caused by the dense HII envelope, the 
masses of objects will be larger and their ages smaller (in 
the diagrams in Fig.~\ref{figure:fig2} this will correspond to 
an upward shift to the left along the absorption line). These 
limiting values are listed in Table~\ref{table:data} taking 
into account the errors of measurements of brightness and 
color indices of the objects. In Table~\ref{table:data} 
we also present galactocentric distances taking into account 
the inclination of the disk $r$ and the diameters of the 
complexes $d$.

Figure~\ref{figure:fig2} shows positions of the complexes 
under study in the color-magnitude diagram and two-color 
diagrams.

\begin{figure}
\vspace{9mm}
\centerline{\includegraphics[width=8cm]{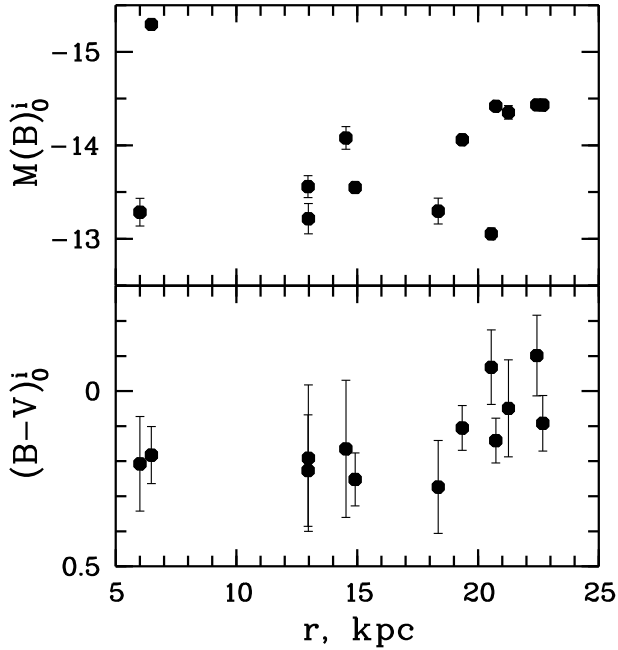}}
\caption{Luminosity and color of stellar complexes depending on their distance to 
the center of the galaxy. Absolute stellar magnitudes $M(B)_0^i$ and color indices 
$(B-V)_0^i$ are corrected for the absorption in the Galaxy and the absorption 
associated with the inclination of the UGC 11973 disk for all complexes, including 
No. 1502. Measurement errors are shown.
\label{figure:fig3}}
\end{figure}

As it can be seen from the figure, in all diagrams, all star 
formation complexes within errors and with account for 
possible underestimate of absorption are located along the 
evolutionary tracks of aging stellar systems. We estimated the 
age and mass of the stellar complex No. 1502, using the 
methodology of \citet{gusev2016}: $t=(2.0\pm1.1)\cdot10^6$~yr, 
$M=(4.6\pm1.6)\cdot10^6 M_{\odot}$. Remaining complexes that 
were not studied by the spectroscopic methods have masses 
$M\ge1.7\cdot10^5 M_{\odot}$ and ages $t\le3\cdot10^8$~yr.

\section{DISCUSSION}

Our estimates of the age of stellar complex no. 1502 indicate 
the time of the last major burst of star formation in it. 
Apparently, the real physical age of the complex formation 
should be large. The time scale of star formation in stellar 
complexes of the size of No. 1502 (550 pc) is about 20 Myr 
\citep{efremov1998}. The estimates of the age of the complex 
obtained by other methods are more sensitive to the presence 
of a relatively ''old'' stellar population and will give 
larger values of $t$. In particular, the method of age 
determination from the equivalent width of the H$\beta$ line 
\citep{copetti1986,stasinska1996,schaerer1998}, calculated 
for a single burst of star formation in the entire region, 
gives for the complex No. 1502 a slightly larger age -- from 
6 to 8 Myr. The reason for this is the superposition of the 
spectrum of the stellar population with the age exceeding 
10 Myr (with a high level in continuum and absorption in the 
H$\beta$ line) over the spectrum of the last burst, which 
reduces the value of EW(H$\beta$).

An indicator of the presence of a young stellar population with 
an age of less than 10 Myr is emission in the H$\alpha$ line. 
Photometry of the galaxy in this line was not made; therefore, 
we cannot confidently state that all studied complexes are the 
regions of HII emission. However, according to the location in 
the two-color diagrams, it can be argued that at least five of 
the bluest complexes (Nos. 1505-1507, 1509, 1510) must be 
younger than 10 Myr (see Table~\ref{table:data} and 
Fig.~\ref{figure:fig2}).

The values of $t$ presented in Table~\ref{table:data} are the 
upper limit of the ages of the studied stellar complexes 
according to their color indices $(U-B)_0^i$ and $(B-V)_0^i$. 
Color index $V-I$, though a less reliable one, indicates even 
younger ages of the complexes: their position in the 
$(B-V)_0^i-(V-I)_0^i$ diagram corresponds to the age 
$t\le5\cdot10^6$~yr for all objects, except the four ones 
with the highest value of $(V-I)_0^i$ (see 
Fig.~\ref{figure:fig2} and Table~\ref{table:data}).

The complex No. 1502 does not stand out among other regions of 
star formation in terms of its observed photometric parameters 
(absolute magnitude and color indices corrected for $A_{\rm G}$ 
and $A_i$, see the position of the lower right ends of the black 
segments in Fig.~\ref{figure:fig2}). Its shift to the upper 
left corners in the diagrams is caused by the large internal 
absorption determined from the spectral data using the Balmer 
decrement. It is rather likely that other stellar complexes in 
UGC 11973 have similar internal absorption values and could be 
located in the diagrams $(B-V)_{\rm c}-M(B)_{\rm c}$, 
$(U-B)_{\rm c}-(B-V)_{\rm c}$ and other ones near the complex 
No. 1502.

In a highly tilted galaxy, such as UGC 11973, selection effects 
play an important role. Absorption caused by the inclination of 
the galactic disk is assumed, in the general case, to be constant. 
Actually, absorption varies along the disk field, it decreases 
from the center to the edge of the galaxy in proportion to the 
decrease in the surface density of the dust in the disk and from 
the far edge of the galaxy to the nearest one along the minor 
axis (see Fig.~\ref{figure:fig1}). Apparently, this is why most 
of the identified stellar complexes are located in the north-west 
part of UGC 11973, closest to us and located in the outer regions 
of the disk at the distances $r>18$~kpc from the galactic center 
(see Table~\ref{table:data}). Due to selection effects, we do not 
analyze the spatial distribution of the complexes in the galaxy.

It should be noted that the complexes closer to the center of the 
galaxy are systematically less bright and more red than the outer 
ones (Fig.~\ref{figure:fig3}). The brightest complex No. 1499, 
which falls out of the general dependence in Fig.~\ref{figure:fig3}, 
has an area many times larger than the other ones (see 
Table~\ref{table:data}).

In general, the measured and estimated parameters of the population 
of stellar complexes in UGC 11973 are typical for star formation 
regions in the large late-type spiral galaxies. A detailed analysis 
of the characteristics of star formation regions in the disks of 
galaxies of various types will be carried out by us in the next 
project work. The study will be based on a homogeneous catalog of 
photometric parameters of more than 1500 star formation regions 
in 19 galaxies \citep[catalog][]{gusev2018}, including the spectral 
characteristics of more than 500 HII regions.

\section{CONCLUSIONS}

1. We have carried out the analysis of the photometric and spectral 
observations of 13 young stellar complexes in the giant spiral 
galaxy UGC 11973. For the complex No. 1502, chemical and physical 
parameters of the surrounding gas, mass and age were estimated.

2. Metallicity of the gas in the vicinity of the stellar complex 
No. 1502 turned out to be, within errors, solar: $Z=0.013\pm0.005$; 
the mass of the complex is $M=(4.6\pm1.6)\cdot10^6 M_{\odot}$, and 
its age $t$ is estimated as $(2.0\pm1.1)\cdot10^6$~yr.

3. For remaining 12 galactic complexes we estimated the lower limit 
of their mass and the upper limit of their age. All complexes 
appeared to be massive $M\ge1.7\cdot10^5 M_{\odot}$, while their 
age does not exceed $3\cdot10^8$~yr.

\section*{FUNDING}

This work was partially supported by the grant of the Moscow State 
University Development Program ''Leading Scientific School: Physics 
of Stars, Relativistic Objects and Galaxies''. This work was carried 
out as part of a scientific project supported since 2020 by the 
Russian Foundation for Basic Research (RFBR) 
(project No.~20-02-00080).

\section*{ACKNOWLEDGEMENTS}

The authors thank the referee for valuable comments and S.N.~Dodonov 
(SAO RAS) for his help in observing with the BTA telescope.

The paper was translated by L.R.~Yungelson.

\end{document}